\documentclass[useAMS,usenatbib,usegraphicx]{mn2e}

\newcommand{\rr}{\tilde{r}}
\renewcommand{\d}{\rmn{d}}
\renewcommand{\O}[1]{\mathcal{O}\!\left[#1\right]}
\newcommand{\eref}[1]{(\ref{eq:#1})}
\newcommand{\fref}[1]{fig. \ref{fig:#1}}
\newcommand{\arxiv}[1]{[arXiv:#1]}
\newcommand{\url}[1]{\texttt{#1}}

\title[Measuring the equation of state of dark matter]
{Combining rotation curves and gravitational lensing:\\
How to measure the equation of state of dark matter in the galactic halo}
\author[T. Faber and M. Visser]{Tristan Faber$^{1}$\thanks{E-mail:
tristan.faber@mcs.vuw.ac.nz} and Matt Visser$^{1}$\thanks{E-mail: matt.visser@mcs.vuw.ac.nz}\\
$^{1}$School of Mathematics, Statistics, and Computer Science, 
Victoria University of Wellington, P.O. Box 600, Wellington, New Zealand}
\begin{document}

\date{8 December 2005; Revised 16 June 2006; \LaTeX-ed \today}

\pagerange{\pageref{firstpage}--\pageref{lastpage}} \pubyear{2006}

\maketitle

\label{firstpage}

\begin{abstract}
  We argue that combined observations of galaxy rotation curves and
  gravitational lensing not only allow the deduction of a galaxy's
  mass profile, but also yield information about the pressure in the
  galactic fluid. We quantify this statement by enhancing the standard
  formalism for rotation curve and lensing measurements to a first
  post-Newtonian approximation. This enhanced formalism is compatible
  with currently employed and established data analysis techniques,
  and can in principle be used to reinterpret existing data in a more
  general context. The resulting density and pressure profiles from
  this new approach can be used to constrain the equation of state of
  the galactic fluid, and therefore might shed new light on the
  persistent question of the nature of dark matter.
\end{abstract}


\section{Introduction}

One of the most compelling issues of modern astrophysics is the open
question concerning the nature of the dark matter which dominates the
gravitational field of individual galaxies and galaxy clusters.  [See
for instance \citet{Persic:1996}, \citet{Borriello:2001}, and
\citet{Salucci:2003}.] While the current consensus in the astrophysics
community is to advocate the cold dark matter (CDM) paradigm, no
\textit{direct} observations of the equation of state have been
carried out to confirm this widely adopted assumption. Efforts to
confirm this assumption include attempts to detect elementary
particles that have been suggested as cold dark matter candidates.
However, experiments that aim (for instance) to detect massive axions
with Earth-based detectors \citep[][\S 22.2.2]{Eidelman:2004} do not
yet yield a positive result.

A different approach to analysing the nature of dark matter has been
suggested by \citet{Bharadwaj:2003} who first proposed that combined
measurements of rotation curves and gravitational lensing could be
used to determine the equation of state of the galactic fluid. Whereas
their analysis made particular assumptions on the form of the rotation
curve, and is restricted to a certain type of equation of state,
herein we provide a general formalism that allows us to deduce the
density and pressure profiles without any prior assumptions about
their shape or the equation of state.

Analytic galaxy halo models that predict a significant amount of
pressure or tension in the dark matter fluid include ``string fluid''
\citep{Soleng:1993}, or some variations of scalar field dark matter
(SFDM). See for instance \citet{Schunck:1999}, \citet*{Matos:1999et},
\citet*{Matos:2000ki}, \citet{Peebles:2000}, and \citet*{Arbey:2003}.
Our method provides a means of observing, or at least constraining,
the pressure distribution in a galactic halo. Therefore it is in
principle able to give evidence for or against specific proposed dark
matter candidates.

The key point is that in general relativity, density and pressure
\textit{both} contribute to generating the gravitational field
\emph{separately}.  Furthermore, the perception of this gravitational
field depends on the velocity of probe particles. These effects become
especially important when one compares rotation curve and
gravitational lensing measurements, where the probe particles are
fundamentally different: interstellar gas or stars at subluminal
velocities for rotation curves, and photons which travel at the speed
of light for lensing measurements. Our formalism accounts for these
crucial differences between the probe particles, and relates
observations of both kinds to the the density and pressure profile of
the host galaxy. Although we (mainly) consider static spherically
symmetric galaxies in a first post-Newtonian approximation, the basic
concept is fundamental and can be extended to more general systems
with less symmetry. A suitable framework for considering most exotic
weak gravity scenarios is provided by the effective refractive index
tensor, as introduced by \citet*{Boonserm:2005}.

The present approach might also help to shed some light on prevailing
problems that arise when combining rotation curve and lensing
observations. For example, an unresolved issue exists when measuring
the Hubble constant from the time delay between gravitationally lensed
images: Using the standard models for matter distribution in the lens
galaxy, the resulting Hubble constant is either too low compared to
its value from other observations, or the dark matter halo must be
excluded from the galaxy model to obtain the commonly accepted value
of $H_0$ \citep{Kochanek:2004}.
A possible explanation of this trend might lie in a disregarded
pressure component of the dark matter halo.

We organise this article in the following manner: First we introduce
the minimal necessary framework of general relativity concepts, and
point out the important conditions required to obtain the Newtonian
gravity limit.  Next, we elaborate on the post-Newtonian extension of
the currently employed rotation curve and gravitational lensing
formalisms.  Consequently, we show how to combine rotation curve and
lensing measurements to make inferences about the density and pressure
profile of the observed galaxy.  
We then examine how noticeable the effects of non-negligible pressure could be in the measurements.
Lastly, we discuss how the formalism adapts to non-spherically symmetric galaxies and comment on the current
observational situation and issues arising with the new formalism.

\section{General relativity framework}

In general relativity the motion of a probe particle is given as the
geodesic of a curved space-time whose curvature is generated by matter
or more generally speaking, stress-energy. The static and
approximately spherically symmetric gravitational field of a galaxy is
represented by the space-time metric \citep*[\S 23.2]{MTW}
\begin{equation}
\label{eq:ss_metric}
\d s^{2}= 
-e^{2\Phi(r)}\, \d t^{2}
+\frac{\d r^{2}}{1-2m(r)/r}
+r^{2}\, \d\Omega^{2} \, ,
\end{equation} 
which is completely determined by the two metric functions $\Phi(r)$
and $m(r)$. These coordinates $(t,r,\theta,\varphi)$ are called
\textit{curvature coordinates} and
\begin{equation}
\d\Omega^{2} = \d\theta^2 + \sin^2\!\theta\:\d\varphi^2
\end{equation} 
represents the geometry of a unit sphere. Invoking the Einstein field
equations with the most general static and spherically symmetric
stress-energy tensor gives the relation between the metric functions
and the density and pressure profiles:
\begin{eqnarray}
\label{eq:Einstein_tt}
8\upi\,\rho(r) \!&=&\! \frac{2\, m'(r)}{r^2} \, ; \\
8\upi\,p_r(r) \!&=&\! - \frac{2}{r^2}\, 
\left[ 
\frac{m(r)}{r} - r\,\Phi'(r)\left( 1-\frac{2\,m(r)}{r} \right)  
\right] ;  \\
\nonumber
8\upi\,p_t(r) \!&=&\! - \frac{1}{r^3}\,\left[ m'(r)\,r-m(r) \right] 
\left[1+r\,\Phi'(r)\right] + \\
\label{eq:Einstein_transverse}
&~& \!\!\!\!\!\!\!\!\!\!\!\!\!\!+\left( 1-\frac{2\,m(r)}{r} \right) 
\left[ \frac{\Phi'(r)}{r} + \Phi'(r)^2 +\Phi''(r) \right] \, ,
\end{eqnarray}
where $\rho(r)$ is the energy-density profile, and $p_r(r)$ and
$p_t(r)$ denote the profiles of the principal pressures in the radial
and transverse directions. Note that we use geometrical units ($c=1$,
$G_\rmn{N}=1$) unless otherwise mentioned.  Hence, if the metric
functions $\Phi(r)$ and $m(r)$ are given by observations, one can
infer the density and pressure profiles. For a perfect fluid, one
would expect $p=p_r=p_t$. From \eref{Einstein_tt} the physical
interpretation of $m(r)$ as the total mass-energy within a sphere of
radius $r$ becomes clear.

\subsection{The Newtonian limit}

Standard Newtonian physics is obtained in the limit of general
relativity where \citep[\S 17.4]{MTW}:
\begin{enumerate}
\item the gravitational field is weak ($2m/r \ll 1$, $2\Phi \ll 1$),
  \label{it:NewtonWeak}
\item the probe particle speeds involved are slow compared to the
  speed of light and \label{it:NewtonSlow}
\item the pressures and matter fluxes are small compared to the
  mass-energy density. \label{it:NewtonNoPressure}
\end{enumerate}

While there is no doubt that in a galaxy condition \ref{it:NewtonWeak}
is satisfied everywhere apart from the central region
\citep{Schoedel:2002}, condition \ref{it:NewtonSlow} only holds for
rotation curves and not for gravitational lensing. Finally, condition
\ref{it:NewtonNoPressure} is related to the open question about the
fundamental nature of dark matter. Hence, the possibility of dark
matter being a high pressure fluid, or some sort of unknown field with
high field tensions, cannot be excluded \textit{a priori}.

It is a standard result that condition \ref{it:NewtonWeak} is enough
to deduce that the gravitational potential $\Phi(r)$ is generated by
the $tt$-component of the Ricci tensor 
\citep{MTW}:
\begin{equation}
\bmath{\nabla}^2 \Phi \approx \mathbfss{R}_{tt} \, ,
\end{equation} 
which on invoking the Einstein equations for $\mathbfss{R}_{tt}$ becomes
\begin{equation}
\label{eq:pressure_field_eq}
\bmath{\nabla}^2 \Phi \approx 4\upi\,(\rho + p_r + 2p_t) \, .
\end{equation}
Therefore, the metric function $\Phi(r)$ can be interpreted as the
Newtonian gravitational potential $\Phi_\rmn{N}$ if and only if the
pressures of the galactic fluid are negligible, i.e. if condition
\ref{it:NewtonNoPressure} holds:
\begin{equation}
\label{eq:Newton_field_eq}
\bmath{\nabla}^2 \Phi_\rmn{N} = 4\upi\,\rho \, .
\end{equation}
It is now quite obvious from \eref{pressure_field_eq} that the
gravitational field is highly sensitive to the pressure if density and
pressure are of the same order of magnitude.

\section{Rotation curves}

For the regime of rotation curve measurements, both conditions
\ref{it:NewtonWeak} and \ref{it:NewtonSlow} apply. In this case, the
geodesic equations of the metric \eref{ss_metric} reduce to
\citep{MTW} 
\begin{equation}
\label{eq:rotcurve_geodesic}
\frac{\d^2 \bmath{r}}{\d t^2} \approx - \bmath{\nabla}\Phi \, ,
\end{equation}
where $\bmath{r}$ denotes the position vector of a probe particle.
Equation \eref{rotcurve_geodesic} is equivalent to the Newtonian
formulation of gravity, except for the general relativistic potential
$\Phi$ which replaces the Newtonian potential $\Phi_\rmn{N}$.

Measurements of rotation curves are carried out by observing the
Doppler shift in the emission lines of the light emitting probe
particles. In a general relativistic context, the observed shift in
wavelength is not exclusively due to Doppler effects of the moving
probe particles, but also depends on the gravitational redshift which
arises as the photons climb out of the gravitational potential well.

It has been shown for edge-on galaxies that the total wavelength
shift\footnote{The wavelength shift that arises from the systemic
  velocity of the galaxy is not considered here, and
  \citet{Faber:2006} has shown that this does not change the result
  presented in this context.} $z_\pm(r)$ of an emission line of a
probe particle at radius $r$ is given by
\citep*{Nucamendi:2000,Lake:2004,Faber:2006}
\begin{equation}
\label{eq:GR_redshift}
1 + z_\pm(r) = \frac{1}{\sqrt{1-r\,\Phi'(r)}} 
\left( 
\frac{1}{e^{\Phi(r)}} - \frac{\pm |b|\,\sqrt{r\,\Phi'(r)}}{r} 
\right) \, ,
\end{equation} 
where prime denotes the derivative with respect to $r$, $' = \d/\d r$
and $b$ is the impact parameter. $z_+$ is the wavelength shift of an
approaching particle and $z_-$ that of a receding particle.

The impact parameter is equivalent to the apparent distance between
the galactic centre and the emitting particle, once one takes notice
of light bending effects. However, in the weak gravity regime of
galaxies, where flat space is a suitable approximation, one finds
\begin{equation}
|b| = r + \O{\Phi} 
\end{equation}
for particles whose position vector $\bmath{r}$ (with respect to the
galactic centre) is perpendicular to the observer's line of sight
\citep{Lake:2004}. Thus, with the additional weak field assumption
$r\,\Phi'(r) \ll 1$, equation \eref{GR_redshift} can be written as
\begin{equation}
1 + z_\pm(r) = 1 \mp \sqrt{r\,\Phi'(r)} + \O{\Phi, r\,\Phi'} \, ,
\end{equation} 
or equivalently,
\begin{equation}
\label{eq:RC_redshift}
z_\pm^2 = r\,\Phi'(r) + \O{\Phi^2, (r\,\Phi')^{3/2}, \Phi \sqrt{r\,\Phi'}} \, .
\end{equation} 
Comparing this expression to the Doppler shift in Newtonian gravity,
\begin{equation}
\frac{v^2}{c^2} = z_N^2 = r\,\Phi_N'(r) \, ,
\end{equation} 
we conclude that for small particle speeds $v \ll c$, i.e. condition
\ref{it:NewtonSlow}, the observation of $z_\pm$ in edge-on galaxies is
in first order equivalent to the Doppler redshift in Newtonian
gravity, when $\Phi_\rmn{N}$ is substituted by $\Phi$. This also
justifies the previous assumption $r\,\Phi'(r) \ll 1$.

For galaxies of arbitrary orientation it is more tedious to obtain
this result, but in a similar fashion, it can also be shown that the
Doppler shift in wavelength is the dominant contribution to the
observed total redshift \citep{Faber:2006}.

Therefore, the usual techniques for obtaining the potential
$\Phi_\rmn{RC}$ from rotation curve measurements can be employed if
one keeps in mind that the motion of the observed particles is not
governed by the Newtonian gravitational potential $\Phi_\rmn{N}$, but
by its general relativistic generalisation $\Phi$:
\begin{equation}
\label{eq:PhiRC}
\Phi_\rmn{RC} = \Phi \neq \Phi_\rmn{N} \, .
\end{equation} 
If one assumes condition \ref{it:NewtonNoPressure}, the density $\rho$
is related to $\Phi_\rmn{RC}$ by \eref{Newton_field_eq}. In the
general case, however, the interpretation of the mass which is
inferred by rotation curve measurements, $m_\rmn{RC}(r)$, can be
obtained from \eref{pressure_field_eq}:
\begin{equation}
\label{eq:pseudomass_rotcurve}
m_\rmn{RC}(r) = r^2\,\Phi_\rmn{RC}' \approx 4\upi\, \int (\rho + p_r + 2p_t)\, r^2\, \d r \, .
\end{equation} 
Therefore, in the general case, we call $m_\rmn{RC}(r) \neq m(r)$ the
\textit{pseudo-mass} determined by rotation curve measurements.

\section{Gravitational Lensing}

A fundamentally different approach of measuring the gravitational
field of a galaxy is gravitational lensing. Here, the observable
photons are not only conveying the information about the gravitational
field to us, they also act as probe particles themselves. Hence,
condition \ref{it:NewtonSlow} is naturally not satisfied for
gravitational lensing observations. Consequently, the equations of
motion for photons \textit{do not} simplify to
\eref{rotcurve_geodesic}, as is the case for rotation curves. Instead,
the geodesic equations for photons have to be solved exactly to
understand the influence of the gravitational field, as it is
described by both metric functions, $\Phi(r)$ \textit{and} $m(r)$.
Fortunately, for certain spacetimes, such as e.g. \eref{ss_metric}, it
is possible to characterise the entire trajectory of light rays with a
single effective refractive index $n(r)$.

\subsection{Fermat's principle and the effective refractive index}

Fermat's principle of shortest optical paths also applies to the
geodesic trajectories of 4-dimensional curved spacetime
\citep{Kline65,MTW}. This description of light rays in a gravitational
field is equivalent to classical optics in a transparent medium with a
continuous refractive index $n$, where Fermat's principle is
formulated as the vanishing of the first variation of the optical
length between two points, $q_1$ and $q_2$, on the trajectory:
\begin{equation}
\delta\!\! \int_{q_1}^{q_2} n(\rr)\,
\left[ \d \rr^{2} + \rr^{2}\, \d\Omega^{2} \right] = 0 \, .
\end{equation} 
By transforming the curvature coordinates of the spacetime
\eref{ss_metric} to so called \textit{isotropic coordinates}
\citep{Perlick:2004},
\begin{equation}
\label{eq:iso_coords}
\d s^{2}= 
e^{2\Phi(\rr)}\, \left\lbrace -\d t^{2}
+n(\rr)^2 \left[ \d \rr^{2}
+\rr^{2}\, \d\Omega^{2} \right] \right\rbrace \, ,
\end{equation}
we introduce the scalar effective refractive index $n(\rr)$ of a
static spherically symmetric gravitational field. By direct comparison
of \eref{ss_metric} and \eref{iso_coords}, we find a differential
equation that relates the $\rr$-coordinate of the isotropic
coordinates to the $r$-coordinate of the curvature coordinates,
\begin{equation}
\label{eq:rr_r}
\frac{\d\rr}{\d r} = \frac{\rr}{r \sqrt{1-2m(r)/r}} \, ,
\end{equation} 
and the refractive index,
\begin{equation}
\label{eq:n_of_rr}
n(\rr) = \frac{r}{\rr}\,e^{-\Phi(r)} \, .
\end{equation}
Since condition \ref{it:NewtonWeak} is satisfied for the region we are
interested in, we can Taylor expand and formally integrate \eref{rr_r}
under appropriate boundary conditions and find
\begin{equation}
\label{eq:rr_of_r}
\rr = r\, \exp \left\lbrace \int \frac{m(r)}{r^2}\, \d r 
+ \O{\left(\frac{2m}{r}\right)^2} \right\rbrace \, ,
\end{equation}
which, inserted into \eref{n_of_rr}, gives
\begin{eqnarray}
n(\rr) \!&=&\! \exp \Bigg\lbrace - \Phi[r(\rr)] - 
\int \frac{m[r(\rr)]}{r(\rr)^2}\, \frac{\d r}{\d\rr}\, \d\rr 
\nonumber\\
&&
\qquad\qquad
+ \O{\left(\frac{2m}{r(\rr)}\right)^2} \Bigg\rbrace ,
\end{eqnarray}
where $r(\rr)$ is given by the inverse of \eref{rr_of_r}. Since $\rr =
r + \O{2m/r}$, the radii in both sets of coordinates are
interchangeable to the desired order and hence, we can also give the
refractive index as a function of the curvature coordinate $r$
directly:
\begin{equation}
\label{eq:refractive_index}
n(r) = 1 - \Phi(r) - \int \frac{m(r)}{r^2}\, \d r + \O{\left(\frac{2m}{r}\right)^2, \frac{2m}{r}\, \Phi, \Phi^2} .
\end{equation}
This effective refractive index entirely determines the trajectory of
a light ray, i.e. the probe particles of gravitational lensing. Hence,
it is the \textit{only} possible observable of gravitational lensing.
We note that the refractive index contains two distinct ingredients,
the potential part, $\Phi(r)$, and the integral over the
mass-function, $\int 2m(r)/r^2\,\d r$.

At this point, we conclude that since gravitational lensing
observations yield $n(r)$ and rotation curve measurements yield
$\Phi(r)$, combined observations of $n(r)$ and $\Phi(r)$ allow the
separate deduction of $\Phi(r)$ and $m(r)$, and therefore describe the
gravitational field of a galaxy in a general relativistic sense,
without any prior assumptions. The fundamental principle is that the
perception of the gravitational field by probe particles depends on
the speed of the probe particles, which manifests itself in the
difference of observables $n(r) \neq \Phi(r)$.

For convenience and comparability, we define the lensing potential as
\begin{equation}
\label{eq:lens_potential}
2\,\Phi_\rmn{lens}(r) = \Phi(r) + \int \frac{m(r)}{r^2}\, \d r \, , 
\end{equation}
so that
\begin{equation}
\label{eq:refindex}
n(r) = 1 - 2\,\Phi_\rmn{lens}(r) + \O{\Phi_\rmn{lens}^2} \, .
\end{equation}

\subsection{Gravitational lensing formalism}

The standard formalism of gravitational lensing in weak gravitational
fields is based on the superposition of the deflection angles of many
infinitesimal point masses \citep*[\S 4.3]{Schneider92}.

In general relativity, a point mass $M$ is described by the
Schwarzschild exterior metric,
\begin{equation}
\d s^{2}= 
-\left(1-\frac{2M}{r}\right)\, \d t^{2}
+\frac{\d r^{2}}{1-2M/r}
+r^{2}\, \d\Omega^{2} \, ,
\end{equation}
which inserted into \eref{refractive_index} gives the effective refractive index
\begin{eqnarray}
\label{eq:refindex_pointmass_decomp}
n(r) &=& 1 + \frac{M}{r} - \int \frac{M}{r^2}\, \d r + \O{\left(\frac{2M}{r}\right)^2} \\
\label{eq:refindex_pointmass}
&=& 1 + \frac{2M}{r} + \O{\left(\frac{2M}{r}\right)^2} \, .
\end{eqnarray} 
In the Newtonian limit, this is generally identified with the Newtonian potential,
\begin{equation}
\label{eq:refindex_Newton}
n(r) = 1 - 2\,\Phi_\rmn{N}(r) \, ,
\end{equation} 
whereas from \eref{refindex_pointmass_decomp} it is clear that in the
general case, the refractive index is only partially specified by the
potential term. That the mass term of the refractive index of a point
mass is identical to the potential term is a special case of the
Schwarzschild metric.  Keeping this in mind, we proceed to outline the
current formalism.

For a point mass, the angular displacement of the gravitationally
lensed image in the lens plane can be calculated from the refractive
index \eref{refindex_pointmass}
\citep{Schneider92}: 
\begin{equation}
\label{eq:pointmass_deflection}
\bmath{\hat\alpha} = 4M\,\frac{\bmath\xi}{|\bmath\xi|^2} \, ,
\end{equation}
where $\bmath\xi$ is the vector that connects the lensed image and the
centre of the lens in the 2-dimensional lens plane. Since the extent
of a lensing galaxy's mass distribution is small compared to the
distance between the light emitting background object and the lens, as
well as compared to the distance between the lensing galaxy and the
observer, one assumes the deflecting mass distribution to be
geometrically thin. Therefore, the volume density $\rho$ of the
lensing galaxy can be projected onto the so-called lens plane,
resulting in the surface density $\Sigma(\bmath\xi)$ which describes
the mass distribution within the lens plane \citep{Schneider:1985}.

The total deflection angle of a lensing mass with finite extent is
then said to be given by the superposition of all small angles
\eref{pointmass_deflection} due to the infinitesimal masses in the
lens plane \citep{Schneider92}:
\begin{equation}
\bmath{\hat\alpha}(\bmath\xi) = 4 \int \Sigma(\bmath\xi')\,
\frac{\bmath\xi-\bmath\xi'}{|\bmath\xi-\bmath\xi'|^2}\, \d^2 \xi' \, .
\end{equation} 
This equation is the foundation of the gravitational lensing formalism
as introduced by \citet{Schneider:1985}, and \citet{Blandford:1986}.

Since this formalism is based on the assumption that the total angle
of deflection is caused by the superposition of point masses --
without the notion of pressure at all -- it automatically assumes that
the underlying lensing potential is Newtonian, i.e. $\Phi_\rmn{lens} =
\Phi_\rmn{N}$. Hence, the lensing potential and the naively inferred
density- or mass-distribution are related by \eref{Newton_field_eq}:
\begin{equation}
\bmath{\nabla}^2 \Phi_\rmn{lens}(r) = 4\upi\,\rho_\rmn{lens}(r) 
\end{equation} 
which implies
\begin{equation}
\Phi_\rmn{lens}(r) = \int \frac{m_\rmn{lens}(r)}{r^2} \, \d r \, .
\end{equation} 
However, as we argued previously, the lensing potential
$\Phi_\rmn{lens}$ is the fundamental observable, and not the density
$\rho$ which was used to construct the formalism. Therefore, for the
general case that does not assume \ref{it:NewtonNoPressure}, i.e.
$\Phi_\rmn{lens} \neq \Phi_\rmn{N}$, we note that
\begin{equation}
\rho_\rmn{lens}(r) \neq \rho(r) 
\quad \rmn{and} \quad 
m_\rmn{lens}(r) \neq m(r) \, .
\end{equation} 
Instead, the deduced mass distribution $m_\rmn{lens}(r)$ has to be
considered as a \textit{pseudo-mass} similar to that of rotation curve
measurements. Its physical interpretation can be deduced from the
definition of the lensing potential \eref{lens_potential}:
\begin{eqnarray}
m_\rmn{lens}(r) &=& \frac{1}{2}\, m_\rmn{RC}(r) + \frac{1}{2}\, m(r) \\
\label{eq:pseudomass_lens}
&\approx&  4\upi\, \int \left[ 
\rho + \frac{1}{2} \left(p_r + 2p_t\right) 
\right] \, r^2\, \d r \, .
\end{eqnarray}

\section{Bringing rotation curves and gravitational lensing together}

We showed in the previous sections that the potentials obtained from
rotation curve and lensing observations, $\Phi_\rmn{RC}$ and
$\Phi_\rmn{lens}$, do not agree in the general case,
\begin{eqnarray}
\label{eq:Phi_rotcurve}
\Phi_\rmn{RC}(r) &=& \Phi(r) \, , \\
\label{eq:Phi_lens}
\Phi_\rmn{lens}(r) &=& \frac{1}{2}\,\Phi(r) + \frac{1}{2}\, \int \frac{m(r)}{r^2}\, \d r \, ,
\end{eqnarray} 
but only in the Newtonian limit, where condition
\ref{it:NewtonNoPressure} holds, in which case $\Phi_\rmn{RC} =
\Phi_\rmn{lens} = \Phi_\rmn{N}$. Since this is the standard assumption
for interpreting rotation curve and lensing data, the results of these
observations are often reported as mass distributions instead of
potentials. Under the Newtonian assumption, the mass and the potential
are related by a field equation of the form \eref{Newton_field_eq}. In
the general case, this leads to the definition of the distinct
pseudo-masses,
\begin{eqnarray}
\label{eq:m_rotcurve}
m_\rmn{RC}(r) &=& r^2\,\Phi'(r) \, ,  \\
\label{eq:m_lens}
m_\rmn{lens}(r) &=& \frac{1}{2}\,r^2\,\Phi'(r) + \frac{1}{2}\, m(r) \, ,
\end{eqnarray} 
which describe the observations equivalently to the potentials
\eref{Phi_rotcurve} and \eref{Phi_lens}. Equations \eref{m_rotcurve}
and \eref{m_lens} can easily be inverted to give the metric functions
$\Phi'(r)$ and $m(r)$,
\begin{eqnarray}
\Phi'(r) &=& \frac{m_\rmn{RC}(r)}{r^2} \, , \\
m(r) &=& 2\,m_\rmn{lens}(r) - m_\rmn{RC}(r) \, ,
\end{eqnarray}
which inserted into the field equations of general relativity
\eref{Einstein_tt}--\eref{Einstein_transverse} yield the density and
pressure profiles:
\begin{eqnarray}
\label{eq:Einstein_approx_tt}
4\upi\,r^2 \rho(r) &=& 2\,m_\rmn{lens}'(r) - m_\rmn{RC}'(r) \, , \\
4\upi\,r^2 p_r(r) &=& 2\, \frac{m_\rmn{RC}(r)-m_\rmn{lens}(r)}{r} 
+ \O{\left( \frac{2\,m}{r} \right)^2} \, , \\
\nonumber
4\upi\,r^2 p_t(r) &=& r\,
\left[ \frac{m_\rmn{RC}(r)-m_\rmn{lens}(r)}{r} \right]' 
 + \O{\left( \frac{2\,m}{r} \right)^2} \\
\label{eq:Einstein_approx_transverse}
&=& \frac{r}{2}\,\left[4\upi\,r^2 p_r(r)\right]'  
+ \O{\left( \frac{2\,m}{r} \right)^2} \, .
\end{eqnarray}
As consistency checks, we note that:
\begin{itemize}
\item The Einstein equations in curvature coordinates,
  \eref{Einstein_approx_tt}--\eref{Einstein_approx_transverse}, agree
  to the given order of $2m/r$ with the Einstein equations of the
  metric in isotropic coordinates, and therefore the approximation
  $\rr \approx r$ is valid;
\item From the equations
  \eref{Einstein_approx_tt}--\eref{Einstein_approx_transverse} follows
  that
\begin{equation}
4\upi\,r^2 \left[ \rho(r) + p_r(r) + 2p_t(r) \right] \approx m_\rmn{RC}'(r) \, ,
\end{equation} 
and
\begin{equation}
4\upi\,r^2 \left[ \rho(r) + \frac{1}{2} 
\left(p_r(r) + 2p_t(r)\right) \right] \approx m_\rmn{lens}'(r) \, ,
\end{equation} 
and thus these results are consistent with the weak field
approximation of the field equations \eref{pressure_field_eq}, and the
interpretations of the pseudo-masses \eref{pseudomass_rotcurve} and
\eref{pseudomass_lens};
\item For $m_\rmn{RC}'(r) = m_\rmn{lens}'(r) = m'(r)$ we find the
  desired result of the Newtonian limit:
\begin{eqnarray}
\label{eq:Newtonlimit_I}
4\upi\,r^2 \rho(r) &=& m'(r) \, , \\
4\upi\,r^2 p_r(r) &=&  \O{\left( \frac{2\,m}{r} \right)^2} \, ,  \\
\label{eq:Newtonlimit_III}
4\upi\,r^2 p_t(r) &=&  \O{\left( \frac{2\,m}{r} \right)^2} \, .
\end{eqnarray}
\end{itemize}
We conclude that the currently existing formalisms for analysing data
from rotation curve and gravitational lensing observations can be used
to separately obtain the pseudo-masses $m_\rmn{RC}(r)$ and
$m_\rmn{lens}(r)$, which by
\eref{Einstein_approx_tt}--\eref{Einstein_approx_transverse} yield the
density and pressure profiles in a first post-Newtonian approximation.
Furthermore, from the combination
\begin{equation}
4\upi\,r^2\,(p_r+2p_t) \approx 2\,(m_\rmn{RC}'-m_\rmn{lens}') \, ,
\end{equation} 
one can immediately infer that the observed system is Newtonian in the
sense of condition \ref{it:NewtonNoPressure} if and only if
$m_\rmn{RC}'(r) \approx m_\rmn{lens}'(r)$. Furthermore, defining the
dimensionless quantity
\begin{equation}
\label{eq:wfactor}
w(r) = \frac{p_r(r)+2p_t(r)}{3\rho(r)} \approx 
\frac{2}{3}\,\frac{m_\rmn{RC}'(r)-m_\rmn{lens}'(r)}{2\,m_\rmn{lens}'(r)-m_\rmn{RC}'(r)}
\end{equation} 
gives a convenient parameter that determines a ``measure'' of the
equation of state.

\section{Parameterizing the size of the effect}
To get an idea how noticeable the existence of a pressure contribution
is likely to be in the measured data, we introduce the $\chi$-factor
which we define as the ratio of the derivatives of $m_\rmn{lens}$ and
$m_\rmn{RC}$:
\begin{equation}
\label{eq:chi_factor}
\chi[w(r)] = \frac{m'_\rmn{lens}(r)}{m'_\rmn{RC}(r)} = \frac{2+3w(r)}{2+6w(r)} \, .
\end{equation}
This can easily be obtained from rearranging \eref{wfactor}.

\begin{figure}
\includegraphics[width=84mm]{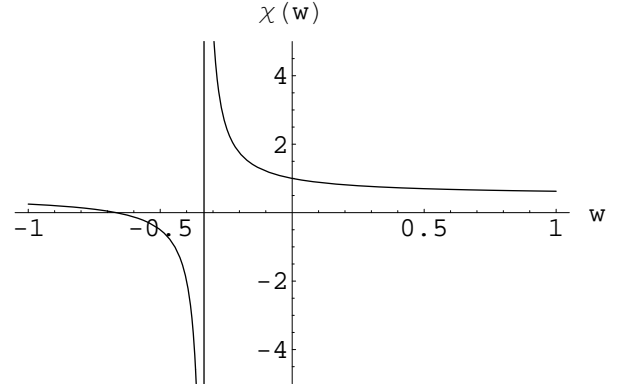}
\caption{The $\chi$-factor \eref{chi_factor} as a function of $w$ in
  the commonly discussed range $w \in [-1,1]$. Naturally $\chi(0)=1$,
  which corresponds to the Newtonian case, see \eref{Newtonlimit_I} --
  \eref{Newtonlimit_III}. At the ends of the plotted range we find
  $\chi(-1)=1/4$ and $\chi(1)=5/8$. The zero-crossing and the
  first-order pole correspond to $w=-2/3$ and $w=-1/3$ respectively,
  as is obvious from \eref{chi_factor}.}
\label{fig:chi_factor}
\end{figure}

Using \fref{chi_factor}, we can now see how the ratio of the slopes of
$m_\rmn{lens}$ and $m_\rmn{RC}$ relates to $w$. One should especially
note that $\chi$ is not very different from unity in the vicinity of
$w=0$, making it difficult to detect small pressures.  However, for $w
\in [1/2, 1]$, which is a range that might plausibly be identified
with real-life data (or at the very least with some real-world
theoretical prejudices), $\chi$ takes values between $62.5\%$ and
$70\%$.  For most negative $w$, specifically for $w$ in the interval
$[-1,0]$, $\chi$ is rather distinct from unity, and therefore should
be easily detectable if the quality of the observational data is high.

Also note that in the special case of a position-independent $w$ ---
so that the ``equation of state of the galactic fluid'' is constant
throughout the observed region --- the ratio between $m_\rmn{lens}$
and $m_\rmn{RC}$ will be the same as between their derivatives:
\begin{equation}
m_\rmn{lens}(r) = \chi(w) \, m_\rmn{RC}(r) \, .
\end{equation}
This relation is likely to be more useful since it does not depend on
numerically obtained derivatives when one wishes to compare mass
profiles. However, it comes at the price of the additional assumption
that $w=\mbox{constant}$.

\section{Non-spherical galaxies}
Although we chose to present this new formalism using the example of a
simple spherically symmetric galaxy, it is easy to show that the
fundamental concept behind the formalism is also valid for
configurations with less symmetry.

\subsection{Rotation curves}
The formalism of Newtonian mechanics is usually adopted when data from
dynamical observations is examined to determine the shape of the
matter distribution of a galaxy. That is, the fundamental equation
employed is \eref{rotcurve_geodesic}:
\begin{equation}
\label{eq:rotcurve_geodesic_no_symm}
\frac{\d^2 \bmath{r}}{\d t^2} \approx - \bmath{\nabla}\Phi(x,y,z) \, .
\end{equation}
Note that no particular symmetry is now assumed for the gravitational
potential $\Phi(x,y,z)$. The only necessary assumptions are the
weakness of the gravitational field [condition \ref{it:NewtonWeak}]
and the slowness of the probe particles [condition
\ref{it:NewtonSlow}].

In view of the formalism presented herein, one now only needs to
realise that the gravitational potential inferred from dynamical
observations is generally not the Newtonian potential
\begin{equation}
\Phi_\rmn{RC}(x,y,z) = \Phi(x,y,z) \neq \Phi_\rmn{N}(x,y,z) \, ,
\end{equation}
and that the valid field equation is [\emph{cf} \eref{pressure_field_eq}]:
\begin{equation}
\label{eq:pressure_field_eq_no_symm}
\bmath{\nabla}^2 \Phi(x,y,z) \approx 
4\upi\,\left[\rho(x,y,z) + \sum\nolimits_{i=1}^3 p_i(x,y,z)\right] \, .
\end{equation}

\subsection{Gravitational lensing}
The notion of a simple effective refractive index can easily be
extended to the class of \emph{conformally static} spacetimes which
are also \emph{conformally Euclidean} \citep{Perlick:2004}. The metric
of this class has no particular spatial symmetry and takes the form:
\begin{equation}
\label{eq:iso_coords_no_symm}
\d s^{2}= 
e^{2\Phi(x,y,z)}\, \left\lbrace -\d t^{2}
+n(x,y,z)^2 \left[ \d x^{2} + \d y^{2} + \d z^2 \right] \right\rbrace \, .
\end{equation}
To account for the weakness of the gravitational field [condition
\ref{it:NewtonWeak}], we assume
\begin{equation}
\Phi(x,y,z) \ll 1 
\end{equation}
and
\begin{equation}
n(x,y,z)=1+h(x,y,z) \quad \mathrm{with} \quad h(x,y,z) \ll 1 \, .
\end{equation}
The first Einstein field equation\footnote{The first Einstein field
  equation is that which is associated with the $tt$-component of the
  Einstein tensor.} is then
\begin{eqnarray}
4\upi \rho(x,y,z) &=& - \bmath{\nabla}^2 \Phi(x,y,z) - \bmath{\nabla}^2 h(x,y,z) 
\nonumber\\
&&
+ \O{\Phi^2, h \Phi, h^2} .
\end{eqnarray}
Inverting this equation yields the effective refractive index
\begin{eqnarray}
n(x,y,z) &=& 1 - \Phi(x,y,z) - 4\upi\, (\bmath{\nabla}^2)^{-1} \rho(x,y,z) 
\nonumber\\
&&
+ \O{\Phi^2, h \Phi, h^2} ,
\end{eqnarray}
where the constants of integration have been chosen to agree with the
special case of spherical symmetry \eref{refractive_index}, and
$(\bmath{\nabla}^2)^{-1}$ is the inverse Laplacian operator. The
general non-spherical lensing potential analogous to
\eref{lens_potential} can be defined as
\begin{equation}
2\,\Phi_\rmn{lens}(x,y,z) = \Phi(x,y,z) + 4\upi\, (\bmath{\nabla}^2)^{-1} \rho(x,y,z) \, ,
\end{equation}
so that the non-spherical refractive index is of the same form as \eref{refindex}:
\begin{equation}
n(x,y,z) = 1 - 2\,\Phi_\rmn{lens}(x,y,z) + \O{\Phi_\rmn{lens}^2} \, .
\end{equation} 
The corresponding field equation for the lensing potential is
\begin{eqnarray}
&& \bmath{\nabla}^2 \Phi_\rmn{lens}(x,y,z) = 
\frac{1}{2} \bmath{\nabla}^2 \Phi(x,y,z) + 2\upi\,\rho(x,y,z) \\
\label{eq:lensing_field_eq_no_symm}
&&= 4\upi\, \left[  \rho(x,y,z) + 
\frac{1}{2}\sum\nolimits_{i=1}^3 p_i(x,y,z)  \right]  \, .
\end{eqnarray}

\subsection{Non-spherical formalism}
Combining the non-spherical field equations
\eref{pressure_field_eq_no_symm} and \eref{lensing_field_eq_no_symm}
yields the density and pressure distributions in absence of any
particular spatial symmetry:
\begin{eqnarray}
4\upi\, \rho \!\! &\approx& \!\!  
2\, \bmath{\nabla}^2 \Phi_\rmn{lens}(x,y,z) - 
\bmath{\nabla}^2 \Phi_\rmn{RC}(x,y,z) 
\\
4\upi\sum\nolimits_{i=1}^3 p_i \!\! &\approx& \!\!  
2 \left[ \bmath{\nabla}^2 \Phi_\rmn{RC}(x,y,z) - 
\bmath{\nabla}^2 \Phi_\rmn{lens}(x,y,z) \right]
\end{eqnarray}
Thus the two observable potentials $\Phi_\rmn{RC}(x,y,z)$ and
$\Phi_\rmn{lens}(x,y,z)$ determine the density and pressure
distributions of a non-spherical galaxy to the lowest order in the
weak gravitational field represented by the functions $\Phi(x,y,z)$
and $h(x,y,z)$.  This is the straightforward extension of the
spherically symmetric formalism presented in this paper.

\section{Observational situation}

The post-Newtonian formalism we have outlined requires the
simultaneous measurement of (pseudo-)density profiles from rotation
curve and gravitational lensing observations.

While in principle these profiles do not have to be of the same
galaxy, they must be comparable in the sense that they accurately
describe ``similar'' galaxies. For example, weak lensing measurements
can be used to statistically infer the (pseudo)-density profile of an
``average'' galaxy \citep{Brainerd:2004}. At the same time, analysing
the dynamics of satellite galaxies gives the rotation curve and thus,
the corresponding pseudo-density profile, of another ``average''
galaxy \citep{Brainerd:2004}. Whether these two ``average'' galaxies
are comparable or not depends on many factors, such as e.g. the
distribution of galaxy morphologies in both samples, the statistical
noise, the employed models for the (pseudo)-density distribution, etc.
These statistical issues render the fast-growing collection of weak
lensing data problematic for our purposes.

On the other hand, combined simultaneous measurements of rotation
curves and lensing of individual galaxies are extremely well suited
for our formalism. However, while there is a large number of
individual rotation curves available \citep[$>
100,000$;][]{Sofue:2001}, the number of individual ``strong'' lensing
systems with multiple images is rather limited\footnote{For an up to
  date list see\\ \url{http://www.cfa.harvard.edu/glensdata/}.}
\citep[$\sim 70$;][]{CASTLE:2005}. Combined observations are further
aggravated by the differing distance scales: Most high quality
rotation curves are naturally available for galaxies with a low to
intermediate redshift of up to $z\sim 0.4$ \citep{Sofue:2001}, while
gravitational lenses are easier to detect at intermediate to high
redshifts \citep[$z \ga 0.4$;][]{CASTLE:2005}, since the image
separation scales increasingly with the redshift of the lensing galaxy
\citep{Schneider:1985,Kochanek:2004}. Therefore, even for nearby
galaxies with existing combined measurements of kinematics and lensing
(e.g. 2237+0305 at $z\approx 0.039$ and ESO 325-G004 at $z\approx
0.035$), the lensing data is restricted to the core region, while the
rotation curve is only described by few data points in the outer
region of the lens galaxy \citep{Barnes:1999,Trott:2002,Smith:2005}.
Consequently, the inferred pseudo-mass profiles are available for the
same galaxy, but unfortunately at different radii and therefore not
comparable.

Although the observational situation makes it currently difficult to
employ the formalism presented, the situation is likely to improve in
the future when observations with a higher resolution will be carried
out -- preferably with an emphasis on obtaining high-resolution
rotation curves for lensing galaxies that exhibit lensed images at
different radii.

\section{Conclusions}

We have argued that the standard formalism of rotation curve
measurements and gravitational lensing make an \textit{a priori}
Newtonian assumption that is based on the CDM paradigm. We introduce a
post-Newtonian formalism that does not rely on such an assumption, and
furthermore allows one to deduce the density- and pressure-profiles in
a general relativistic framework. In this framework, rotation curve
measurements provide a pseudo-mass profile $m_\rmn{RC}(r)$ and
gravitational lensing observations yield a different pseudo-mass
profile $m_\rmn{lens}(r)$. Combining both pseudo-masses allows one to
draw conclusions about the density- and pressure
profiles\footnote{These formulae are given in SI units, hence the
  factor of $c^2$.} in the lensing galaxy,
\begin{eqnarray}
\rho(r) &=& \frac{1}{4\upi\,r^2} \, 
\left[ 2\,m_\rmn{lens}'(r) - m_\rmn{RC}'(r) \right] \, , \\
p_r(r)+2p_t(r) &\approx& \frac{2\,c^2}{4\upi\,r^2} \, 
\left[m_\rmn{RC}'(r)-m_\rmn{lens}'(r) \right] \, . 
\end{eqnarray}
In the case of absent or negligible pressure, this could be used to
observationally confirm the CDM paradigm of a pressureless galactic
fluid. Conversely, if significant pressure is detected, a
decomposition of the galaxy morphology would allow new insight into
the equation of state of dark matter.

For instance, detailed observation of the recently discovered closest
known strong lensing galaxy ESO 325-G004 \citep{Smith:2005} could
provide satisfactory data to allow the decomposition of density and
pressure of the galactic fluid, as outlined in this article. The
system consists of an isolated lensing galaxy at redshift $z\approx
0.035$ with an effective radius of $R_\rmn{eff}=12\,\farcs 5$ and
arc-shaped images of the background object at $R\approx 3\arcsec$, and
possible arc candidates at $R\approx 9\arcsec$. \citet{Smith:2005}
intend to collect more detailed data that hopefully will include
extended stellar dynamics and hence, allow for a direct comparison of
the rotation curve and lensing data, if the arc candidates at
$R\approx 9\arcsec$ turn out to contribute to the measurements.

Since the formalism presented is based on a first-order weak field
approximation, we suggest that to confirm the findings, one should
re-insert the obtained density and pressure profiles into the metric
\eref{ss_metric}. The actual observed quantities can then be extracted
numerically for comparison from the exact field equations
\eref{Einstein_tt}--\eref{Einstein_transverse} and the geodesic
equations.

Finally, even though data might not yet be available to constrain the
dark matter equation of state noticeably, one should note that the
possibility of non-negligible pressure in the galactic fluid
introduces a new free parameter into the analysis of combined rotation
curve and lensing observations.

\section*{Acknowledgments}

We thank Silke Weinfurtner for some helpful suggestions and comments.
This research was supported by the Marsden Fund administered by the
Royal Society of New Zealand (MV), the J.L. Stewart Scholarship, and a
Victoria University of Wellington Postgraduate Scholarship for
Master's Study (TF).


\bsp

\label{lastpage}

\end{document}